\documentclass[twocolumn,showpacs,preprintnumbers,prl,aps,psfig]{revtex4}
\usepackage{graphicx}
\usepackage{dcolumn}
\usepackage{amsmath}
\usepackage{latexsym}
\usepackage{longtable}
\begin{document}

\title{Electrical resistance of Ni nanowires (diameter $\geq$ 20 nm) near the Curie Temperatures}
\author{M. Venkata Kamalakar and A. K. Raychaudhuri \\}
\affiliation{DST Unit for Nanosciences, Department of Material Sciences, S N Bose National Centre for Basic Sciences, Block-JD, Sector-III, Salt Lake, Kolkata-700098, India.\\}
\date{\today}

\begin{abstract}
In this letter we report electrical transport measurements on nickel nanowires of diameters  down to 20 nm in the region close to its paramagnetic-ferromagnetic transition temperature $T_C$ (reduced temperature$|t| \leq 10^{-2}$). The data analysis done in the frame work of critical behavior of resistance near $T_C$  shows that the critical behavior persists even down to the lowest diameter wire of 20 nm.  However, there is a suppression of the critical behavior of the resistivity as measured by the critical exponents and the parameters quantifying  the anomaly. The spin system shows approach to  a quasi one-dimensional spin system.

\end{abstract}
\maketitle
Resistivity ($\rho$) of a metallic nanowire is a topic of considerable current interest. In addition to effects such as Quantum transport, even in the regime of classical transport there are a number of effects that have received scant attention. One such phenomenon is the issue of electrical resistance near the Curie temperature($T_C$) for paramagnetic-ferromagnetic transition. In this range, the issues associated with Critical point phenomena become important \cite {stanley}. While in bulk ferromagnets and to some extent in thin films of ferromagnetic materials like Ni,Co and Fe, the anomaly in resistivity near the transition temperature $T_C$ has been well studied and analyzed in terms of critical  phenomena \cite{CP1,CP2}, there exists no such study in ferromagentic nanowires whose diameters  can become smaller than the spin correlation length.  One important issue that becomes crucial when the size is reduced  is finite size scaling, which can describe the shift of the ferromagnetic $T_C$ as a fucntion of size \cite{Fisher_barber}. In the context of ferromagnetic nanowires, this particular point (shift of $T_C$ with size) has been looked into in only one study through determination of $T_C$ from magnetization measurements \cite{Sun}. However, issues like critical exponents have not been investigated. In this paper we report an extensive investigation of the anomaly associated with the resistivity near the  $T_C$,  through an electrical transport measurement (Resistance($R$) vs. Temperature($T$)) in the temperarure range $300 K \leq T\leq 675 K$ in Ni nanowires of diameter ranging from $200$ nm to $20$ nm. We note that, to the best of our knowledge, such resistance measurements upto this temperature range have not been done in metal nanowires before. This measurement, made close to   the critical temperature $T_C$ ($|t| \leq 10^{-2}$, where $t=\frac{T-T_{c}}{T_{c}}$), allows us to investigate whether the critical anomaly in $\rho$ persists even in nanowires with diameters less than the correlation length. This investigation is of significance in view of the recent observation in Zn nanowires with sizes much less than the superconducting coherence length, the specific heat anomaly near the transition temperature is essentially unaltered and there is not much "rounding-off" of the critical region \cite{Kurtz}.  

Resistivity anomaly near a continuous transition has been associated with the specific heat anomaly($C$) \cite{Fisher_Langer} so that $\frac{1}{\rho}\frac{d\rho}{dT} \propto C$. This has been proved in precision measurements on Ni \cite{CP1,CP2}. The critical exponent $\alpha$ determined from both specific heat as well as resistivity anomaly were found to be similar in bulk Ni. There is no theory or experiment to prove the validity of the Fischer-Langer theory in the region of nanowires where the effects of finite size should be visible. However, we note that the premises of the  theory depends on short range spin-correlation which we believe may be sufficiently unaltered in these wires and thus there may be reason enough to assume the validity of the above theory. (Note: Our analysis of the results, which we present below, however, in no way depend on the validity of the above relation).

In this work, we report measurements of resistance ($R$) of nickel nanowires of various diameters (20 nm, 55 nm, 100 nm, 200 nm) in the temperature region $300 K-675 K$ which encompasses the region $T\sim T_C$, where the critical behavior  shows up. The $R$ of a 50 $\mu$m diameter nickel wire (99.999\% purity) is also measured  as the reference bulk wire. The synthesis of nickel nanowires is done by electrodeposition of nickel from a 1M NiCl$_{2}$.7H$_{2}$0 solution into the pores of nanoporous Anodic Alumina Membranes (AAM) having  thickness in the range 50 $\mu$m-60 $\mu$m. The diameters of pores inside the membranes have a very narrow size distribution (rms value $< \pm 5\%$) and the wires grown in them have nearly the same diameter as the pore. The details of the method of preparation has been given elsewhere \cite{venkat1}. X-ray diffraction (XRD) of the nickel filled membranes revealed fcc structure of the nickel nanowire arrays with lattice constant of 3.54 $\AA$. Energy dispersive spectroscopy further confirmed the purity of the nanowires. A Transmission electron micrograph and HRTEM of 20 nm diameter nanowires are shown in Fig.1a and Fig.1b respectively. To improve the homogeneity and to reduce the resistance due to grain boundary contribution, the nanowires were annealed at 670 K for 24hrs in a vacuum of 10$^{-6}$ mbar. XRD pattern of the Ni nanowires after vaccuum annealing did not show any peaks of NiO ruling out any oxidation of the samples. The annealing increased the average  grain size to a size approximately similar to their diameters and also made the wire stable and the resistance ($R$) decreased to a low value. (Note: The average grain size  as well as the microstrain were estimated from the XRD line widths  using the Williamson Hall plot \cite{Williamson}).

\begin{figure}[t]
\centering
\includegraphics[width=8cm,height=4cm]{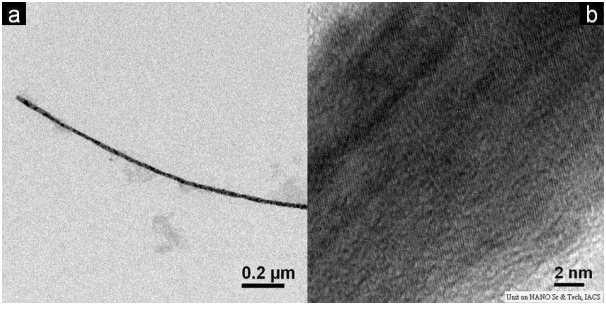}
\caption{(a) TEM image of a 20 nm diameter nickel nanowire.
(b) HRTEM image of an edge of a 20 nm diameter nickel nanowire.}
\end{figure}

The resistance of the bulk nickel wire  (50 $\mu$m) was measured by four probe method while for  the nanowire arrays  a pseudo-four probe technique \cite{Bid} was used by retaining the wires inside the templates as shown in the inset of figure Fig. 2a. The leads were attached using high temperature silver epoxy. For measurement,  a phase-sensitive detection technique using a low frequency  AC signal was used and the samples were kept in a high vacuum (10$^{-6}$ mbar) in a specially designed high temperature resistivity measurement setup. The temperature was measured and  controlled by a temperature controller (with PT100 sensor) that allows   resolution and control to a $mK$. Close to T$_{C}$, the resistance of the samples were measured at a slow ramp of 0.1 K/min. Fig. 2a shows the $R$ data of the nanowire arrays (normalized with the value of $R$ at 300 K) as compared with $R$ of the bulk. The $T_C$ as determined by quantitative analysis, described later on, are marked by arrows.

\begin{figure}[t]
\centering
\includegraphics[width=7cm,height=10cm]{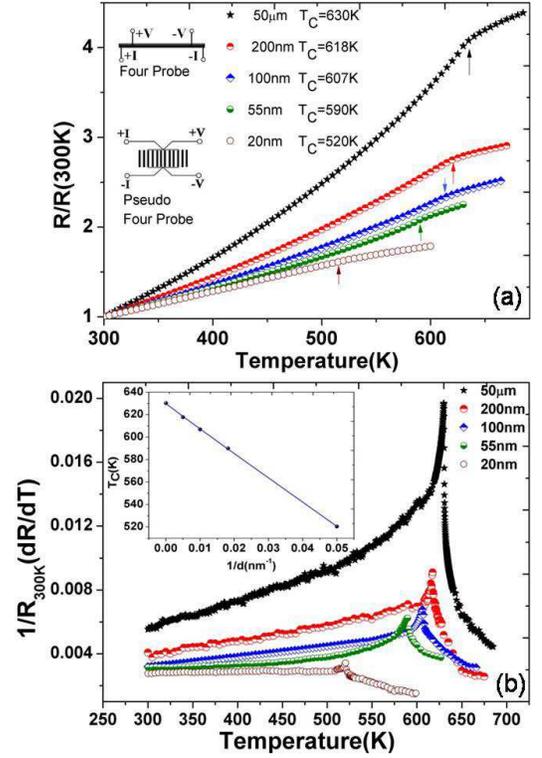}
\caption{(a)Normalized plot of $R$ vs. $T$ for nickel nanowires of varying diameters. The T$_{C}$ as determined from the resistance anomaly are shown by arrows. (b) Plot of  ($\frac{1}{R_{300K}}\frac {dR}{dT}$ vs. $T$) for nickel nanowires. The inset shows the variation of $T_C$ with $d^{-1}$.}
\end{figure}

 First we present some of the basic observations. In Fig. 2a, it can be seen that the anomaly in the resistance reduces as the diameter is reduced (along with a distinct reduction in $T_C$). However, the anomaly, though reduced is present even in the wire with the smallest diameter. To accentuate the anomaly in $R$ we show in Fig. 3 the derivative  of the resistance ($\frac{1}{R_{300K}}\frac {dR}{dT}$) as a function of $T$. The presence of the distinct feature in the derivative near $T_C$  is present in all the wires, though much reduced in the wires of smaller diameter. The dependence of  T$_{C}$  on wire diameter is shown as an inset in  Fig. 2b plotted as $T_C$ vs.$1/d$. The reduction in $T_C$ (as determined from the resistance  data)  with diameter $d$ is very similar to that obtained from magnetization measurements \cite{Sun}. The dependence of $T_C$ can be described by the finite-size scaling laws \cite{Fisher_barber}. The finite size effect can be  observed as a change of T$_C$ when the correlation length $\xi$(T) is compareble to $d$. The asymptotic behavior of the correlation length $\xi$(T) of a magnetic system close to the bulk transition temperature is described by \cite{Fisher_barber} :
\begin{equation}
\xi(T)=\xi_{0}|t|^{-\nu}
\end{equation}

where $\xi_{0}$ is the extrapolated value of $\xi$(T) at T=0 K and  $t=\frac{T-T_{c}}{T_{c}}$ is the reduced temperature. The critical temperature T$_{C}$(d) scales with the diameter of the nanowires as :
\begin{equation}
\frac{T_{C}(\infty)-T_{C}(d)}{T_{C}(\infty)} ={\left ( \frac{\xi_{0}}{d}\right )}^{\lambda}
\end{equation}

where $\lambda = 1/\nu$.
 We obtain  $\lambda$=0.98. The value obtained from the results of resistivity anomaly agrees very well with those obtained from magnetization measurements($\lambda$=0.94) for nanowires \cite{Sun}. However it is lower than the values predicted by 3D Heisenberg Model ($\lambda$=1.4) and 3D Ising Model( $\lambda$=1.58) \cite{stanley,schneider}. The value of $\xi_{0}$ (=3.35 nm), is larger than but comparable to that obtained from the magnetization measurements \cite{Sun}.

We now investigate the crucial issue of whether the resistance anomaly near $T_C$ can indeed be analyzed using the frame work of critical phenomena. In nanowires such issues like strain as well as other effects can lead to a rounding-off of the transition and we would have to address this issue when analyzing the data in the  critical region. The resistivity data near $T_C$ in bulk Ni wires have been adequately analyzed by precision experiments in two earlier reports and there are well accepted values for the critical exponents \cite{CP1,CP2}. To standardize our analysis we use the data on the $50 \mu$m sample and analyze it using the protocol as described earlier \cite{CP1,CP2},  in the region around $T_C$ for $|t| \leq 10^{-2}$.
We use the equation:
\begin{equation}
R(t)=R_{0}+R't+\left\{
\begin{array}{rl}
A_{-}|t|^{(1-\alpha)}(1+D_{-}|t|^z),& (t<0);\\
A_{+} t^{(1-\alpha)}(1+D_{+} t^z), & ( t\geq 0).
\end{array} \right.
\end{equation}
\\
 The terms $R_{0}$ and $R' t$ mainly consist of the non-critical part of the resistance of the sample. The third term is the critical part of the resistance responsible for the power law behavior. $\alpha$ is the critical exponent with $A_{-}$ and $A_{+}$ being the critical amplitudes of the leading term. We note that in the frame work of the Fisher-Langer theory this $\alpha$ is the same as the specific heat exponent. ${Z}$ is the exponent and $D_{-}$ and $D_{+}$ are the amplitudes of the second order correction to scaling or the confluent singularity. ${Z=0.55}$ as predicted by the renormalization group theory \cite{Le}. Though the Eq.(3) represents the behaviour of resistance near T${_C}$ for a homogeneous material with no spread in T${_C}$, one can correct for the inhomogeneties by using a convolution of the form: \cite{CP1,CP2}

\begin{equation}
R^{*}(T,T_{C},\sigma)=\int R(T, T_{C}-x)g_{\sigma}(x)dx
\end{equation}
where $g_{\sigma}(x)$ is a Gaussian in $x$ of width $\sigma$ and $x$ is the temperature variance from an average $T_{C}$. The integration was numerically solved using a 32 point Gauss-Hermite quadrature and a least square fitting was done with variable parameters $R_{0}$, $R'$ , $\alpha$, $A_{-}$,  $A_{+}$, $D_{-}$,  $D_{+}$, ${T_{c}}$,  $\sigma$  with a maximum fit error of 0.04\%. A typical fit for the bulk wire is shown in Fig. 3 with the fitting error shown in the top inset. In the region very close to $T_C$ there is a deviation due to rounding-off which is discussed later on. The extracted parameters are tabulated in Table.I.  along with the exponents obtained from the previous work on Ni. The comparison shows that the analysis used by us is standardized and thus can be used to analyze the data for the nanowires.
\begin{figure}[t]
\centering
\includegraphics[width=6.5cm,height=5cm]{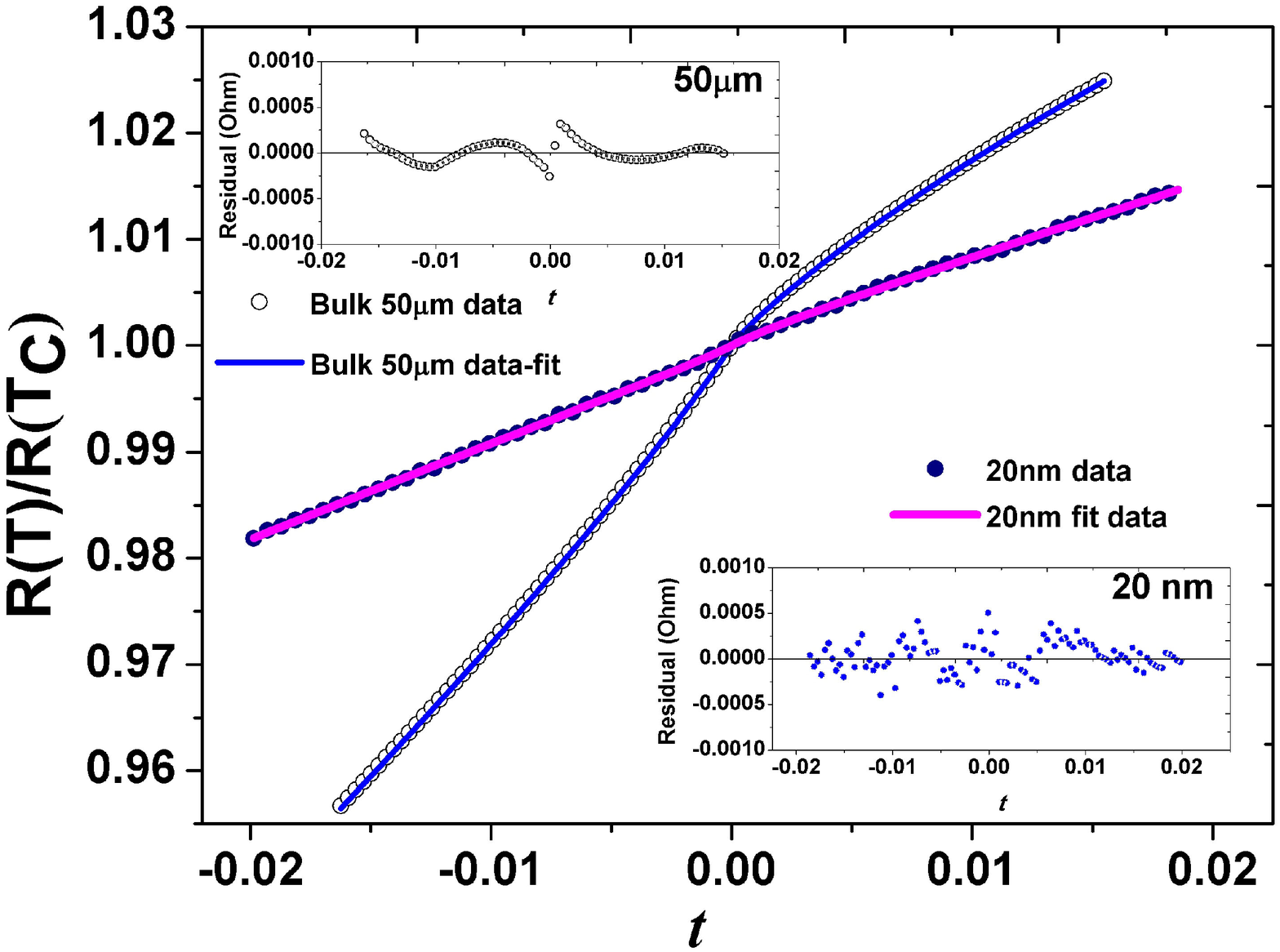}
\caption{$\frac{R(T)}{R(T_C)}$ vs. $t$ (alongwith the fits) for the 20 nm and bulk wire about T$_{C}$($t=0$). The insets show the residuals for bulk and 20 nm diameter nanowire data fits.}
\end{figure}

In the context of nanowires, the first point that one would like to investigate is whether there is a critical behavior and whether the existing frame work of analysis at least operationally describes the behavior of the resistance near $T_C$. We do not put forward any theoretical arguments whether such analysis is justifiable but instead investigate empirically whether such approach can be used. The exponent $\alpha$ in Eq. (3) which describes the critical behavior is the same as the exponent $\alpha$ that describes the critical behavior of heat  capacity. Since we do not know whether this is valid in such systems of reduced dimension we use the exponent $\alpha_r$ instead. In nanowires $\alpha_r$ may or may not be the same as $\alpha$, the heat capacity exponent.

\begin{table*}
\caption{\label{tab:table1} The exponents obtained for the bulk and the nanowires by fit as compared with previous data.}
\begin{ruledtabular}
\begin{tabular}{ccccccc}
 Reference/Diameter(nm) & T$_{C}$(K)& $\alpha_r$ & $A_{+}/A_{-}$ &  $D_{+}/D_{-}$ & $\sigma$  \\
\hline
Bulk Ref. 2& 630.284 $\pm$0.003 &-0.095 $\pm$0.002 &-1.53 $\pm$0.02&-0.8 $\pm$0.1&0.13$\pm$0.12\\
Bulk Ref. 3& 630.650 $\pm$0.450&-0.089$\pm$0.003 &-1.48$\pm$0.03 &-1.2$\pm$0.2& 1.28$\pm$0.03\\
50000& 630.248$\pm$0.001 &-0.09957$\pm$0.00004 &-1.506$\pm$0.001 &-0.981$\pm$0.031&0.1409$\pm$0.0069\\
200& 617.752$\pm$0.003 &-0.08638$\pm$0.00063 & -1.423$\pm$0.013 &-0.845$\pm$0.028&0.4349$\pm$0.0247\\
100&606.903$\pm$0.017 & -0.08123$\pm$0.00104 &-1.382$\pm$0.021&-0.737$\pm$0.148&0.3214$\pm$0.0013\\
55& 590.004$\pm$0.008 &-0.06381$\pm$0.00051&-1.290$\pm$0.012 &-0.103$\pm$0.022&0.4447$\pm$0.0147\\
20& 520.339$\pm$0.011 &-0.03086$\pm$0.00011 & -1.003$\pm$0.002 & 0.466$\pm$0.048&0.5214$\pm$0.0158\\

\end{tabular}
\end{ruledtabular}
\end{table*}
\noindent

Fit of Eq. (4) to the data for 20 nm diameter nanowire is also shown in Fig. 3. The data can be fitted to  Eq.(4) with reasonable accuracy which is comparable to that obtained for the bulk data. The exponents and other parameters obtained show a very clear trend as the  diameter is progressively reduced (see Table-I). In case of the nanowires, there is a considerable decrease in the magnitude of $\alpha_r$ showing a slowing down of the growth of the $dR/dT$ near the critical point. In addition, changes in $A_{+}/A_{-}$  and $D_{+}/D_{-}$  are also prominent.  The variations in the values of the  $\alpha_r$, amplitude ratios and the correction terms are plotted in Fig. 4. The above analysis establishes two points: (1) The critical behavior as manifested in the resistance anomaly is present even in the smallest diameter nanowire (although suppressed) and it is describable within the mathematical frame work given in Eq.(4) and (2) the amplitude of the critical behavior and also the correction to scaling  are  severely suppressed as the diameter is reduced. 

\begin{figure}
\centering
\includegraphics[width=6.5cm,height=5cm]{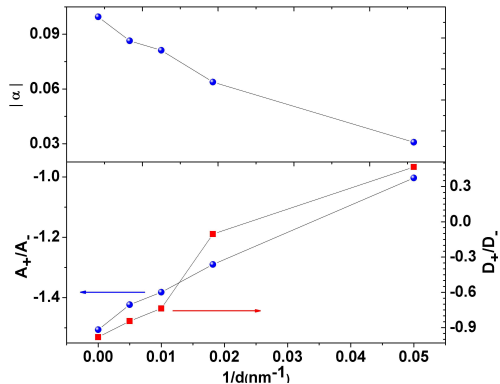}
\caption{ Variation $|\alpha|$ and amplitude ratios $A_{+}/A_{-}$ and $D_{+}/D_{-}$ for the wires studied as determined from the scaling relation (Eq. (4)).}
\end{figure}

While analyzing the data near the critical point the rounding off becomes an important issue. This is present even in bulk systems like Ni and in nanowires, in particular, when we do the measurements in an array of nanowires, these can be severe. To avoid the effect of rounding-off in the estimation of the exponents often the small regions in temperatures ($|t| \approx 10^{-3}$) around the $T_C$ is left out in data analysis. We have estimated the rounding-off region below and find that it is limited to a region ($\approx \pm 2 K$) even in the smallest diameter wire. We have redone the analysis leaving out the data in the region $T_C \pm 2 K$ around $T_C$ and find that the exponent $\alpha_r$ does not change much.

The observed spread in $T_C$ as measured by the width of $\sigma$ of the Gaussian in Eq.(4) is a good empirical measure of the minimum rounding off that we may expect. In the bulk sample we used $\sigma\approx 0.14 K$ which is comparable to that seen in most bulk samples before (see Table I). In the nanowires, irrespective of the size the typical $\sigma\approx 0.4-0.5 K$. 
The spread in $T_C$ and the resulting  rounding-off in the resistance anomaly in our measurements arise from the following two principal sources: (1) Spread due to strain inhomogeneity, (2) Spread in the wire diameters as the measurement is done in an array. We can estimate them and compare that with experiments. From the X-ray micro-strain measurements which give the strain inhomogeneity in the wires we can estimate the expected spread in $T_C$ due to strain inhomogeneity using the bulk modulus of Ni and the dependence of $T_C$ on pressure($\frac{dT_{C}}{dP}$) \cite{CP2}. We find that for all the nanowires the expected maximum spread in $T_C$ is in the range of $0.4 -0.7 K$ which is of the same magnitude as $\sigma$ that one obtains from the fit to the resistance data. The expected spread in $T_C$ from the distribution in size can be calculated from the measured distribution and the finite size scaling relation (Eq.2). The estimated spread in $T_C$ is $\leq 0.9 K$ for the larger diameter wires and is $\leq 2 K$ for the smallest diameter wire. Thus it appears that most of the contribution causing spread in $T_C$ comes from the wire size distribution.

The above analysis establishes that the resistivity anomaly exists in ferromagnetic nanowires at their transition temperature $T_C$. The anomaly and the associated critical behavior is progressively but systematically suppressed (reduction in absolute values of the ratio $A_{+}/A_{-}$ as well as $\alpha$ as the diameter of the wire is reduced). The correction to scaling as quantified by the ratio $D_{+}/D_{-}$ shows a systematic change from negative to positive sign.

The effect of finite size on the critical exponents should be visible in the temperature and size regime we are working. As the transition temperature is approached, the correlation length gets constrained by the wire diameter. Hence the truncation of the divergence of the correlation length begins as the temperature approaches T$_{C}$. This happens at $|t| \leq 0.17$ in the case of 20 nm wire and $|t| \leq 0.036$ in the case of 100 nm wire. This is the range of $|t|$ where we observe critical behaviour. In case of strictly one dimensional nanowires, there should not be a second order transition if one considers the 1D Ising model \cite{Gitterrnan}. Mathematically speaking, the critical part of $R$ has to vanish or become a constant. This implies the ratio $A_{+}/A_{-} \rightarrow -1 $ and $\alpha \rightarrow 0$ with D$_-$ and $D_+$ becoming insignificant. We observed  $\alpha_r \rightarrow 0$ and also the observed ratio $A_{+}/A_{-} \rightarrow -1 $. This would imply that the spin system is tending towards 1-D behavior. We note again , as we have pointed out before, we are not sure whether in such systems as ours we will have $\alpha$ (from heat capacity) same as $\alpha_r$. However, we also note that in one class of systems with restricted diameter where the value of $\alpha$ from specific heat experiment has been evaluated rigorously is in superfluid transition of He4 in pores with diameter in the same range.  It has been seen that in these 1D constrained systems the value of $\alpha$ is -0.02 \cite{Gasparini} which is comparable with the value we obtained ($\approx$ -0.03 in 20 nm wires).

In summary, these are the first measurements of the resistance of nickel nanowires of various diameters in the critical region ($T_C\pm 10 K$), where one would expect anomaly in resistivity due to critical point phenomena. With the decrease in diameter, we observed a decrease in the transition temperature. The data analysis shows that the observed resistance data shows critical behaviour down to the lowest diameter wire of 20 nm  and the anomaly close to $T_C$ can be treated in the frame work of resistance behavior near a critical region. However, there is a suppression of the critical behavior of the resistivity including a decrease in the magnitude of  $\alpha$ and the ratio$A_{+}/A_{-} \rightarrow -1 $ as one would expect in a quasi one-dimensional spin system. 

The authors thank the Department of Science and Technology, Govt. of India and CSIR, Govt. of India for financial support in the form of a Unit and sponsored scheme. MVK acknowledges CSIR, Govt. of India for fellowship. DST Unit for nanosciece and technology, IACS, Kolkata is acknowledged for providing TEM support.

\end{document}